\documentclass[12pt]{article}

\usepackage{amsmath,amssymb,amsfonts}
\usepackage[paper=letterpaper,margin=0.9in]{geometry}

\newcommand{\be}{\begin{equation}}
\newcommand{\ee}{\end{equation}}
\newcommand{\bea}{\begin{eqnarray}}
\newcommand{\eea}{\end{eqnarray}}
\newcommand{\ba}{\begin{array}}
\newcommand{\ea}{\end{array}}
\newcommand{\eref}[1]{(\ref{#1})}

\newcommand{\BC}{{\mathbb C}}

\newcommand{\CP}{{\mathbb C}{\mathbb P}}

\newcommand{\Diff}{{\rm Diff}}
\newcommand{\Gr}{{\rm Gr}}

\newcommand{\tr}[1]{{\rm tr}\, {1}}
\newcommand{\bra}[1]{\langle{1}|}
\newcommand{\ket}[1]{|{1}\rangle}
\newcommand{\ip}[2]{\langle{1}|{2}\rangle}
\newcommand{\vev}[1]{\langle{1}\rangle}

\newcommand{\todo}[1]{{\bf {1}}}
\newcommand{\comment}[1]{}

\def\dd{{\rm d}}

\newcommand{\ft}[2]{{\textstyle\frac{1}{2}}}
\def\fract12{{\textstyle{1\over2}}}
\def\ffract12{\raise .3 em\hbox{$\scriptstyle1$}\kern-.25em/
                \kern-.2em\lower .2 em \hbox{$\scriptstyle2$}}
\def\fractje12{{\scriptstyle{1\over2}}}

\def\part12{{\partial1\over\partial2}}
\def\ex1{e^{\textstyle1}}

\begin{document}

\renewcommand{\thepage}{\arabic{page}}
\setcounter{page}{1}


\vskip 2.00 cm
\renewcommand{\thefootnote}{\fnsymbol{footnote}}

\centerline{\bf \Large  Modeling Time's Arrow}
\vskip 0.75 cm

\centerline{{\bf
Vishnu Jejjala,${}^{1}$\footnote{\tt vishnu@neo.phys.wits.ac.za}
Michael Kavic,${}^{2}$\footnote{\tt michael.kavic@liu.edu}
Djordje Minic,${}^{3}$\footnote{\tt dminic@vt.edu} and 
Chia-Hsiung Tze${}^{3}$\footnote{\tt kahong@vt.edu}
}}
\vskip .5cm
\centerline{${}^1$\it School of Physics}
\centerline{\it University of the Witwatersrand}
\centerline{\it WITS 2050, Johannesburg, South Africa}
\vskip .5cm
\centerline{${}^2$\it Department of Physics}
\centerline{\it Long Island University}
\centerline{\it Brooklyn, New York 11201, U.S.A. }
\vskip .5cm
\centerline{${}^3$\it Department of Physics}
\centerline{\it Virginia Tech}
\centerline{\it Blacksburg, VA 24061, U.S.A.}
\vskip .5cm
\setcounter{footnote}{0}
\renewcommand{\thefootnote}{\arabic{footnote}}

\begin{abstract}
Quantum gravity, the initial low entropy state of the Universe, and the problem of time are interlocking puzzles. In this article, we address the origin of the arrow of time from a cosmological perspective motivated by a novel approach to quantum gravitation. Our proposal is based on a quantum counterpart of the equivalence principle, a general covariance of the dynamical phase space. We discuss how the nonlinear dynamics of such a system provides a natural description for cosmological evolution in the early Universe. We also underscore connections between the proposed non-perturbative quantum gravity model and fundamental questions in non-equilibrium statistical physics.
\end{abstract}

\setcounter{footnote}{0}
\renewcommand{\thefootnote}{\arabic{footnote}}

\newpage

\section{Introduction}

The second law of thermodynamics, perhaps the deepest truth in all of science, tells us that the entropy is a non-decreasing function of time.
The arrow of time implies that the Universe initially inhabited a low entropy state and the subsequent cosmological evolution took the Universe away from this state.
A theory of quantum gravity must explain the singular nature of the initial conditions for the Universe.
Such a theory should then, in turn, shed light on time, its microscopic and directional nature, its arrow.
The quest for a theory of quantum gravity is fundamentally an attempt to reconcile two disparate notions of time.
On the one hand, Einstein's theory of general relativity teaches us that time is ultimately an illusion.
On the other hand, quantum theory tells us that time evolution is an essential part of Nature.
The failure to resolve the conflict between these competing notions is at the heart of our inability to properly describe the earliest moment of our Universe, the Big Bang.
The puzzle of the origin of the Universe intimately connects to a second fundamental issue:
should the initial conditions be treated separately from or in conjunction with the basic framework of the description of the dynamics?
The discord between the general relativistic and quantum theoretic points of view has very deep roots.
In this essay we will trace these roots to their origins.
Then, drawing inspiration from the profound lessons learned from relativity and quantum theory, we propose a radical yet conservative solution to the problem of time.

In quantum mechanics, time manifests as the fundamental evolution parameter of the underlying unitary group.
We have a state $|\psi\rangle$, and we evolve it as $e^{-\frac{i}{\hbar} H t} |\psi\rangle$.
The Hamiltonian operator of a given system generates translations of the initial state in time.
Unlike other conjugate quantities in the theory such as momentum and position, the relation between time and energy, which is the observable associated to the Hamiltonian, is distinguished.
Time is not an observable in quantum theory in the sense that generally there is no associated ``clock'' operator.
In the Schr\"odinger equation time simply enters as a parameter.
This conception of time as a Newtonian construct that is global or absolute in a post-Newtonian theory persists even when we promote quantum mechanics to relativistic quantum field~theory.

In contrast, time in general relativity is local as well as dynamical.
Suppose we promote general relativity to a quantum theory of gravity in a na\"{\i}ve fashion.
In the path integral, the metric of spacetime is one more dynamical variable.
It fluctuates quantum mechanically.
So notions such as whether two events are spacelike separated become increasingly fuzzy as the fluctuations amplify.
Indeed, Lorentzian metrics exist for almost all pairs of points on a spacetime manifold such that the metric distance is not spacelike.
Clearly the notion of time, even locally, becomes problematic in quantum gravitational regimes.
The commutation relation $[{\cal O}(x),{\cal O}(y)] = 0$ when $x$ and $y$ are spacelike separated, but this is ambiguous once the metric is allowed to fluctuate.
The failure of microcausality means that the intuitions and techniques of quantum field theory must be dramatically revised in any putative theory of quantum~gravity.

Crafting a theory of quantum gravity that resolves the problem of time is a monumental undertaking.
From the previous discussion, we see that the standard conceptions of time in quantum theory and in classical general relativity are in extreme tension:
time in the quantum theory is an absolute evolution parameter along the real line whereas in general relativity there can be no such one parameter evolution.
A global timelike Killing direction may not even exist.
If the vacuum energy density is the cosmological constant $\Lambda$, we may inhabit such a spacetime, de Sitter space.

In any attempt to reconcile the identity of time in gravitation and canonical quantum theory, one is also immediately struck by the remarkable difference in the most commonly used formulations of the two theories.
Whereas general relativity is articulated in a geometric language, quantum mechanics is most commonly thought of in algebraic terms within an operatorial, complex Hilbert space formalism.
A principal obstacle to overcome rests with the r\^ole of time being intrinsically tied to the underlying structure of quantum theory, a foundation which---as recapitulated below---is rigidly fixed.
Yet when quantum theory is examined in its less familiar geometric form, it mimics general relativity in essential aspects.
In fact, these parallels provide a natural way to graft gravity into the theory at the root quantum level.
Particularly, the geometric formulation illuminates the intrinsically statistical nature and rigidity of time in quantum theory and points to a very specific way to make time more elastic as is the case in general relativity.
Thus, by loosening the standard quantum framework minutely, we can surprisingly deduce profound implications for quantum gravity, such as a resolution of the problems of time and of its arrow.
In subsequent sections, we will lay out a framework for general quantum relativity in which the geometry of the quantum is identified with quantum gravity and time is given a dynamical statistical interpretation.
Some bonuses stem from this point of view.
We achieve a new conceptual understanding of the origin of the Universe, the unification of initial conditions, and a new dynamical framework in which to explore these and other issues. 
Here to reach a broader audience, conceptual rendition takes precedence over mathematical formalism the details of which are available in the literature given below.

In Section~2, we examine the problem of time's arrow.
In Section~3, we briefly recall the geometric formulation of standard quantum mechanics, which naturally leads to a generalized background independent quantum theory of gravity and matter, the latter of which is embodied by M(atrix) theory.
In Section~4, we apply this prescription to a theory of quantum gravity in its cosmological setting:
the initial low entropy state of the Universe and cosmological evolution away from this state.
Specifically we will discuss key properties of the new space of quantum states---a nonlinear Grassmannian---features which notably embody an initial cosmological state with zero entropy and provides a description of cosmological evolution when viewed as of a far from equilibrium dissipative system.
We also will elaborate on a more general connection between quantum gravity, the concept of holography and some fundamental results in non-equilibrium statistical physics. Finally, Section~5 offers some concluding~remarks.

\section{The Problem of Time's Arrow}

The basic problem of time is exemplified by the fundamental clash between the classical general relativistic notion of spacetime---through the r\^ole of the diffeomorphism group and the ultimately chimerical nature of the evolution parameter---and the quantum notion of time as the fundamental evolution parameter of the underlying unitary group.
In thinking about the nature of time in physics it is tempting to succumb to the allure of general relativity and declare time to be an illusion and seek a reformulation of quantum theory to fit Einstein's worldview.
The consequent emergence of actual dynamics from the constrained description of such a ``quantum'' theory of Einstein gravity thus presents the main and almost insurmountable problem.

Given the overwhelming success of local quantum field theory in accounting for all (albeit non-gravitational) quantum interactions, it is tempting to retain its underlying precepts as we refine and deepen our analysis.
Thus, Euclidean local quantum field theory was extended into the realm of quantum gravity and in particular to give a prescription for the resolution of the initial singularity or the reconciliation of the problem of the initial conditions as opposed to the fundamental dynamical prescription.

Yet the problem of time runs much deeper; for there is an inescapable puzzle here.
The observational evidence points to a hot Big Bang some $13.75\pm 0.11$ Gyr ago~\cite{wmap}.
The Penrose--Hawking theorems point to the existence of an initial singularity~\cite{ph}.
Under conservative assumptions, there exist geodesics that are only finitely extendable into the past.
Time has a beginning.
The curvature of spacetime blows up at the initial singularity, which is the terminus of the geodesics, and general relativity can no longer reliably say anything useful about physics since the very notion of spacetime geometry breaks down.

The origin of the Universe also raises conceptual questions within the quantum theory itself.
Quantum mechanics tells us that we can evolve a state forward and backward in time forever.
But if we uphold the canonical definition of the unitary evolution with a globally defined evolution parameter in a cosmological setting, what happens to the evolution as we approach the singularity?
According to quantum theory, the evolution is not supposed to terminate, and yet, from both the physics and the general relativity standpoints, it seemingly does end.
Thus arise questions as to the quantum embodiments of the Big Bang, of the initial and final singularities of gravitational collapse, of the nature of the quantum vacuum in such extreme regimes, of the fuzziness of space and time, indeed the question of the very notion of the quantum itself.
Does the quantum need to be transgressed?

A proposed resolution to some of the above issues may be sought in an analysis of time in standard quantum theory set in a differential geometric form.
Clearly, before tackling the issue of its direction and its arrow, we should first settle on what time is in the context of quantum theory.
We begin by noting the difference between space and time in non-relativistic quantum mechanics.
No insight is gained through the usual operatorial formalism where quantum theory looks like a set of phenomenally successful but ultimately {\it ad hoc} set of rules for computation.
Particularly, there is no time operator neither in the point particle, nor even in the string and M-theory settings which leave unresolved the problem of time or equivalently that of background independence.
(AdS/CFT~\cite{adscft, adscft2, adscft3}, which is undeniably a background independent formulation of string theory in anti-de Sitter space, nevertheless fixes the asymptopia.)
From the particle physics or string theory perspective, time is needed for the dynamics while space could be holographically emergent~\cite{holo, holo2, seiberg}. In other words, time may well be more fundamental than space. Subsequently we will take this hunch seriously.

Guided by these ruminations, a more illuminating picture of the nature of quantum theory and of time is to be found through a geometric formulation of quantum theory.
This more intuitive form of quantum theory suggests a way to extend in the quantum context the concept of global time to local dynamical time and provides in the same vein the quantum geometrical framework to accommodate a quantum theory of gravity from first principles.
Such a frontal approach bypasses the ambiguous if not dubious process of canonical quantization of general relativity which, after all, is an effective theory of gravitation.
  In fact, the thermodynamic formulation of classical gravity~\cite{thermo, thermo2, thermo3, ted} already suggests that gravitation is an intrinsically quantum phenomenon, universal in that it couples to all forms of energy.

\section{Quantum Gravity and a New Take on Time}

\subsection{Geometric Quantum Theory}

To begin addressing the problem of time through the lens of quantum gravity, one is struck by the dramatic difference in the formalisms standardly used to describe the two theories.
To recapitulate, general relativity is typically considered in geometric terms while quantum theory is most commonly couched in the operatorial formalism.
As has so often been the case in the history of physics, challenging questions have been difficult to answer in large part because they are cast in a formalism ill suited to providing clear answers.
Some examples are:
(a) Schr\"odinger's formulation of quantum mechanics versus Heisenberg's in easily obtaining the spectrum of the hydrogen atom;
(b) the second quantized formulation of the quantum many body problem as applied to solids and liquids leading, say, to the BCS theory of superconductivity;
(c) the Feynman path integral formulation and its elucidation of perturbative and specially non-perturbative phenomena in relativistic quantum systems.
In such cases, when the correct formalism was found solutions to previously intractable problems became apparent and even natural.
It is in this spirit that we take our first step in resolving the problem of time by considering an alternative but equivalent formulation of quantum mechanics in purely geometric terms.
This process will allow us to do  a direct comparison, when attempting to understand the different ways in which time is treated in general relativity and quantum mechanics.

It is not widely known that standard quantum mechanics may be cast geometrically as Hamiltonian dynamics over a specific phase space $\CP^n$, the complex projective Hilbert space of pure quantum states~\cite{geoqm, geoqm2, as}.
Deferring to the sizable literature for greater details---see~\cite{review} and references therein---we only sum up here the defining features of this geometric formulation.
The state space $\CP^n$ is a compact, homogeneous, isotropic, and simply connected K\"ahler--Einstein manifold with a constant, holomorphic sectional curvature $2/\hbar$.
Notably, being K\"ahler, it possesses a defining property: a triad of compatible structures, any two of which determine the third.
These are a symplectic two-form $\omega$, a unique Fubini--Study metric $g$, and a complex structure $j$, respectively.
Indeed all of the key features of quantum mechanics are in fact encoded in this geometric structure.
In particular, the Riemannian metric determines the distance between states on the phase space and encodes the probabilistic structure of quantum theory. The Schr\"odinger equation is simply the associated geodesic equation for a particle moving on $\CP^n = U(n+1)/(U(n)\times U(1))$ in the presence of an effective external gauge field (namely, the $U(n)\times U(1)$ valued curvature two-form) whose source is the Hamiltonian of a given physical system.
We observe that $\CP^n$, where generally $n\ = \infty$, remains an absolute background in that there is no backreaction from the wavefunction.
So it provides only a single rigid kinematical stage for quantum dynamics.
Another simple but telling feature is the following.
When the configuration space of the theory is three dimensional physical space itself, the Fubini--Study metric reduces to the flat spatial metric.
This observation suggests that space and, indeed, curved spacetime, need not be inputs but may emerge from an, albeit,  suitably extended quantum theory over phase space, generalized both kinematically and dynamically.
To put our results in their proper context, we next summarize the pertinent features of such a generalized geometric quantum theory.

\subsection{Background Independence and Matrix Theory}

First, we recall that crucially, Matrix theory is a manifestly second quantized, non-perturbative formulation of M-theory on a fixed spacetime background~\cite{bfss}.
While physical space emerges as a moduli space of the supersymmetric matrix quantum mechanics, time still appears as in any other canonical quantum theory.

Time is not an observable in quantum mechanics:
as we have repeatedly emphasized,
there is generally no ``clock'' operator in the way there is a position operator or a momentum operator.
Moreover, as we demand diffeomorphism invariance in a theory of gravitation, time and spatial position are only labels, and when the metric is allowed to fluctuate, classical notions, such as spacetime paths and spacelike separation of points---hence causality---cease to have operational meaning.
To construct a background independent
formulation of Matrix theory, it becomes necessary to relax the very stringent rigidity of the underlying quantum theory. Hence the question:
How can this procedure be minimally accomplished?

Our extension of geometric quantum mechanics via a quantum equivalence principle yields the following~\cite{review, mt, mt2, dj, dj2, jm, jm2,footnote1}.
At the basic level, there are only dynamical correlations between quantum events.
Staying close to the structure of quantum mechanics, the phase space of states must have a symplectic structure, namely a symplectic two-form, and be the base space of a $U(1)$ bundle. Moreover---and this is the key difference with unitary quantum mechanics---this  space must be diffeomorphism invariant.
Paralelling quantum mechanics, we next demand a three-way interlocking of the Riemannian, the symplectic, and the non-integrable almost complex structures.
In thus departing from the integrable complex structure of $\CP^n$, which in fact is the {\it only} change made to the structure of quantum mechanical state space, the most natural quantum mechanical phase space is then identified as the nonlinear Grassmannian,
$\Gr(\BC^{n+1}) = \Diff(\BC^{n+1})/\Diff(\BC^{n+1}, \BC^n\times \{0\})$,
with $n\to\infty$, a complex projective, strictly almost K\"ahler manifold.
Here the term nonlinear denotes the fact that we have a coset space of diffeomorphism groups as compared to the cosets from such linear Lie groups as $U(n)$ or $SU(n)$---$\CP^n$ being a pertinent example.
This shift toward a novel state space is then the radical yet conservative departure from standard quantum mechanics.
From diffeomorphism invariance, which defines a new framework beyond quantum mechanics, it follows that not just the metric but also the almost complex structure and hence the symplectic structure be fully dynamical in the space of states.
Consequently, with the coadjoint orbit nature of $\Gr(\BC^{n+1})$, the equations of motion of this general theory are the Einstein--Yang--Mills equations:
\be
\label{BIQM1}
{\cal{R}}_{ab} - \frac{1}{2} {\cal{G}}_{ab} {\cal{R}}  - \lambda {\cal{G}}_{ab}= {\cal{T}}_{ab} (H, F_{ab}) ~
\ee %
with ${\cal{T}}_{ab}$ as determined by ${\cal{F}}_{ab}$, the holonomic Yang--Mills field strength, the Hamiltonian (``charge'') $H$, and a ``cosmological'' term $\lambda$.
Furthermore,
\be
\label{BIQM2}
\nabla_a {\cal{F}}^{ab} = \frac{1}{2M_P} H u^b ~
\ee
where $u^b$ are the velocities, $M_P$ is the Planck energy, and $H$ the Matrix theory Hamiltonian~\cite{bfss}.
These coupled equations imply via the Bianchi identity a conserved energy-momentum tensor: $\nabla_a {\cal{T}}^{ab} =~0$.
Just as the geodesic equation for a non-Abelian charged particle is contained in the classical Einstein--Yang--Mills equations, so is the corresponding geometric, covariant Schr\"odinger equation.
The latter is here genuinely nonlinear and cannot be, as in quantum mechanics, linearized by lifting to a flat Hilbert space.
The above set of equations defines the physical system (here the model Universe) and identifies the correct variables including time.
Just as the geometry of $\CP^n$ holds the key to all the quantum propositions, the crux of our extended quantum theory rests in the above Grassmannian.
We next list its key geometrical features from which new physics is to be deduced.

\subsection{Geometry of $\Gr(\BC^{n+1})$}

In contrast with $\CP^n$, the space $\Gr(\BC^{n+1})$ is much less studied. We do know that it a compact, homogeneous but non-symmetric, multiply-connected, infinite dimensional complex Riemannian space.
It is a projective, strictly almost K\"ahler manifold, a coadjoint orbit, hence a symplectic coset space of the volume preserving diffeomorphism group~\cite{hv}.
It is also the base manifold of a circle bundle over $\Gr(\BC^{n+1})$, where the $U(1)$ holonomy provides a Berry phase.

Essential for our purposes, nonlinear Grassmannians are Fr\'echet spaces.
As generalizations of Banach and Hilbert spaces, Fr\'echet spaces are locally convex and complete topological vector spaces.
(Typical examples are spaces of infinitely differentiable functions encountered in functional analysis.)
Defined either through a translationally invariant metric or by a countable family of semi-norms, the lack of a true norm makes their topological structures more complicated.
The metric, not the norm, defines the topology.
Moreover, there is generally no natural notion of distance between two points so that many different metrics may induce the same topology.
In sharp contrast to $\CP^n$ with its unique Fubini--Study metric, the allowed metrical structures are much richer and more elastic, thereby allowing novel probabilistic and dynamical applications.
Thus $\Gr(\BC^{n+1})$ has in principle an infinite number of metrics, a subset of which form the solution set to the Einstein--Yang--Mills plus Matrix model equations we associate with the space.
For example, in~\cite{mm}, an infinite one-parameter family of non-zero geodesic distance metrics was found.

Since $\Gr(\BC^{n+1})$ is the diffeomorphism invariant counterpart of $\CP^n$, the necessary topological metric to consider is the analogue of the Fubini--Study metric of standard quantum theory.
This weak metric was analyzed in 2005 by Michor and Mumford~\cite{mm}, who obtained a most striking result, henceforth called their {\em vanishing theorem}.
Their  theorem states that the generalized Fubini--Study metric induces on $\Gr(\BC^{n+1})$ a vanishing geodesic distance.
Such a paradoxical phenomenon is due to the curvatures being unbounded and positive in certain directions causing the space to curl up so tightly on itself that the infimum of path lengths between any two points collapses to zero.
What could be the possible physics of such a remarkable property?

\section{Modeling Time's Arrow}

\subsection{The Big Bang}

In order to address the origin of the cosmological arrow of time we must have an understanding of the physics of the initial low entropy state of the Universe.
To this end, we shall apply the quantum gravitational model outlined in the previous section to address the Big Bang singularity.
Remarkably, given the vanishing geodesic distance induced on $\Gr(\BC^{n+1})$ by the generalized Fubini--Study metric, such a state arises as a feature of the model considered here.

Recalling that all of the key features of quantum mechanics are encoded in $\CP^n$, our next crucial task is to take seriously the unusual mathematical properties of $\Gr(\BC^{n+1})$ and interpret them in physical terms.
By taking this space as the space of states out of which spacetime emerges, we see that the vanishing theorem naturally describes an initial state in which the Universe exists at a single point, the cosmological singularity.

Moreover, viewed through this lens, a statistical notion of time may apply close to the cosmological singularity.
To see the intrinsically probabilistic nature of time, we observe that in both standard geometric quantum mechanics and its extension given above, the Riemannian structure encodes the statistical structure of the theory.
The geodesic distance is a measure of change in the system, for example through Hamiltonian time evolution.
In standard quantum mechanics, by way of the Fubini--Study metric and the energy dispersion $\Delta E\ $, the infinitesimal distance in phase space is
\be
\dd s = \frac{2}{\hbar}\, \Delta E\, \dd t ~
\label{distance}
\ee
Through this Aharonov-Anandan relation, time reveals its fundamental statistical, quantum nature \cite{geoqm}.
It also suggests that dynamics in time relate to the behavior of the metric on the state space.
A quantum state changes infinitesimally from fluctuation to fluctuation: {e.g.}, time evolution corresponds to statistical energy fluctuations.
Note also the above linear relation Equation~\eref{distance} between the line element $\dd s$ and the time interval $\dd t$.
It underscores the rigidity of global time in quantum theory, a rigidity connected to the uniqueness and universality of the Fubini--Study metric on $\CP^n$, singled out by the compatibility between symplectic structure and the integrable complex structure.
It is in fact the relaxation of the complex structure to a non-integrable almost complex structure that leads to a more flexible local time and naturally to the need to go beyond the unitary group of quantum mechanics to the diffeomorphism group as the invariance group of the space of states. So in lieu of Equation~\eref{distance} and due to the availability of the Planck length in our model, we can define a local time $t$ through the equation $\dd s =\frac{2}{\hbar}E_p \dd t$ where $E_p$ is the Planck energy \cite{jm2}. In this fashion the time $t$ is only obtained by solving for a given metric of the nonlinear Grassmannian.
Thus, this quantum time in the extended theory is not only local but dynamical;
this is in line with general relativity: here the equivalence principle operates on the quantum state space and not in spacetime seen as an emergent structure yet to be obtained.

Moreover as Wootters~\cite{woot} showed, what the geodesic distance $\dd s$ on $\CP^n$ measures is the optimal distinguishability of nearby pure states:
if the states are hard to resolve experimentally, then they are close to each other in the metrical sense.
Statistical distance is therefore completely fixed by the size of fluctuations.
A telling measure of the uncertainty between two neighboring states or points in the state space is given by computing the volume of a spherical ball $B$ of radius $r$ as $r\to0$ around a point $p$ of a $d$-dimensional manifold ${\cal M}$.
This is given by
\be
\frac{\textrm{Vol}( B_p(r) )}{\textrm{Vol}( B_e(1) )} = r^d \left( 1 - \frac{R(p)}{6(d+1)} r^2  + o(r^2) \right) ~
\label{balls}
\ee
where the left hand side is normalized by $\textrm{Vol}( B_e(1) )$, the volume of the $d$-dimensional unit sphere.
$R(p)$, the scalar curvature of ${\cal M}$ at $p$, can be interpreted as the average statistical uncertainty of any point $p$ in the state space~\cite{petz}.
As $2/\hbar$ is the sectional curvature of $\CP^n$, $\hbar$ can be seen as the mean measure of quantum fluctuations.
Equation~\eref{balls} indicates that, depending on the signs and values of the curvature, the metric distance gets enlarged or shortened and may even vanish.

In fact the vanishing geodesic distance under the weak Fubini--Study metric on $\Gr(\BC^{n+1})$ is completely an effect of extremely high curvatures~\cite{mm}.
Because the space is extremely folded onto itself, any two points are indistinguishable ({\em i.e.}, the distance between them is zero).
This feature is an exceptional locus in the Fr\'echet space of all metrics on $\Gr(\BC^{n+1})$.
It is a purely infinite dimensional phenomenon, and one that does not occur with the $\CP^n$ of the canonical quantum theory.

From the foregoing discussion on quantum local time and the phase space metric, the low entropy problem tied to the initial conditions of the Universe finds a natural resolution.
In the language of statistical or information geometry and quantum distinguishability, the generalized Fubini--Study metric having vanishing geodesic distance between any two of its points means that none of the states of our nonlinear Grassmannian phase space can be differentiated from one another.
Due to the large fluctuations in curvatures everywhere, the whole phase space is comprised of a single state,  an {\em unique} microstate at time zero.
The number of microstates $\Omega$ = 1, so the probability $P$ for this state is unity.
By Boltzmann's thermodynamic entropy formula $S = k_B \log \Omega = - k_B \sum_i P_i \log P_i$---the latter equation being for the informational (Shannon) entropy---it follows that the entropy of the Universe is identically zero at the Big Bang.
We must bear in mind that our model for quantum cosmology begins with the space of states from which space-time subsequently emerges. Thus the vanishing theorem refers to the vanishing of microstates. This implies that the number with the natural weak Fubini--Study metric, the Universe is in that one fixed configuration of just one microstate with probability one. Clearly this exceptional state at time zero begs the question as to the origin of the expected manifold microstates which must occur as the Universe evolves away from the Big Bang. This issue relates directly to the process of nonlinear generation of further degrees of freedom through entropy production.
As we shall see below, from the physics standpoint this striking zero-entropy phenomenon is most natural when viewed within a non-equilibrium setting.

From the relation between geodesic distance and time, we also have the emergence of a cosmological arrow of time.
While at time $t = 0$ the system has entropy $S=0$, the very high curvatures in $\Gr(\BC^{n+1})$ signal a non-equilibrium condition of dynamical instability.
Because of its nonlinear dissipative and chaotic dynamics, our system will flow through small perturbations toward differentiation. This process of entropy production by way of fractionization of one state into ever more distinguishable states in the state space which thus acquires an ever larger volume.
Mathematically, this instability is further evidenced by the existence of a whole family of non-zero geodesic distance metrics, of which the zero entropy metric is a special case~\cite{mm}.
Larger geodesic distance is naturally associated with distinguishably of states and thus larger entropy. Thus evolution from the zero distance state to the family of non-zero geodesics distance metrics allows for a clear example of time evolution (and cosmological evolution) in the direction of higher entropy states.
 In accord with the second law of thermodynamics, the dynamical evolution---made possible, say, via these very metrics---is toward some higher entropy but stable state.
The system will ultimately go into a phase with a dynamically realized lower symmetry in which classical spacetime and canonical quantum mechanics would emerge.
However, it is the physics of the Big Bang---at time zero and immediately afterwards---which will be detailed next.

\subsection{Cosmological Evolution from a Jammed State}

What could the physics behind a low (zero) entropy, yet high temperature state of the Big Bang be?
We suggest that the paradoxical zero distance, everywhere high curvature property of $\Gr(\BC^{n+1})$ with the Fubini--Study metric finds an equally paradoxical physical realization in the context of our model.
This kind of occurrence is to be found in a class of far from equilibrium collective phase transitions, the so called ``freezing by heating'' transitions.

From many studies~\cite{pettini} it has been established that high curvatures in the phase or configuration manifold of a physical system precisely reflect large fluctuations of the relevant physical observables at a phase transition point.
This correspondence means equating the high curvatures of the Fubini--Study metric on $\Gr(\BC^{n+1})$ with large fluctuations in our system at a phase transition.
The vanishing geodesic distance can be interpreted as the signature, or order parameter, of a strong fluctuation (or ``heat'') induced zero entropy, and hence a highly ordered state.

While from an equilibrium physics perspective such a state seems nonsensical, it occurs in certain far from equilibrium environments.
Specifically, we point to a representative continuum model~\cite{helbing, helbing2} where such an unexpected state was first discovered.
Here, one has a system of particles interacting, not only through frictional forces and short range repulsive forces, but also and most importantly via strong driving fluctuations ({e.g.}, noise, heat, \emph{etc.}).
As the amplitude of the fluctuations ({e.g.}, temperature) goes from weak to strong to extremely strong and as its total energy increases, such a system shows a thermodynamically counterintuitive evolution from a fluid to a solid and then only to a gas.
At and beyond the onset of strong fluctuations, it first goes to a highly ordered, low entropy, indeed a crystalline state, which is a phase transition like state if both particle number and fluctuations are sufficiently large.
This jammed collective state, being energetically metastable then goes into a third disordered, higher entropy gaseous state under extremely strong fluctuations.
So from this non-equilibrium setting, the zero entropy property of the Big Bang is compatible with an early Universe observed to have a very nearly uniform matter distribution in thermal equilibrium at a uniform temperature.

While our model's dynamics are clearly mathematically far more intricate than the above models for phenomena such as traffic jams and the flocking of birds, it does have the requisite combination of the proper kind of forces to achieve these ``freezing by heating'' transitions.
The system being considered is far from equilibrium with low entropy, high temperature, and negative specific heat.
In addition we have nonlinear, attractive, and repulsive Yang--Mills forces, short range repulsive forces of D$0$-branes in the Matrix theory, repulsive forces from a positive ``cosmological'' term, and most importantly large gravitational fluctuations induced by the large curvatures.
Moreover, it is known that geometric quantum mechanics can be seen as a classical Hamiltonian system, one with a K\"ahler phase space.
Its complete integrability in the classical sense~\cite{block} derives from the K\"ahler property which implies hermiticity of all observables in their operatorial representations.
The extended quantum theory is similarly viewed in terms of classical nonlinear field and particle dynamics over a strictly almost complex phase space.
This last property implies that corresponding operators are non-hermitian, and hence our system is a dissipative system~\cite{rajeev}.
Moreover, classical Einstein--Yang--Mills systems are non-integrable and chaotic~\cite{barrow}.

Further physical insights into the above quantum gravity model must await the detailed mathematical analysis yet to be done on the space $\Gr(\BC^{n+1})$. This is the subject of the still little explored area of geometric nonlinear functional analysis.
In the meanwhile, we may wish to approach the jammed Big Bang state, its evolving entropy, and its possible decay modes from a more phenomenological and quantifiable stance of non-equilibrium physics of jammed systems. Here we may conceive the following scenario:  Immediately following the Big Bang phase transition, spacetime is expected to emerge as a jammed state of a spacetime foam or quantum foam. It is known that if condensed matter materials such as foams or emulsions are compressed, they display solid behavior above the so-called ``jamming'' transition. We do not wish to commit ourselves to any particular model of spacetime foam discussed in the literature. After all a more specific spacetime foam model or family of models may ultimately derive from the proposed theory of quantum gravity discussed here.  In lieu of making use of a specific model of spacetime foam we here invoke an effective theory of jamming which best describes the universal features of jammed systems and hence possibly jammed spacetime foams.  Edwards' statistical mechanics of jammed matter~\cite{makse} far from equilibrium is precisely such a theory \cite{footnote2}.
So if the above identified Big Bang state is modeled as a granular system with a well defined volume (so a relevant variable), but where, due to dissipation, energy is of minor relevance, then as this state becomes unjammed, the number of microstates $\Sigma_{\rm jammed}(V)$ for a given volume $V$ is equal to the area of the surface ${\cal{W}}(\kappa)=V$ in phase space.
Thus,
\be
\Sigma_{\rm jammed}(V) = \int d\kappa\ \delta\big(V-{\cal{W}}(\kappa )\big)\Theta(\kappa) ~
\ee

This type of system may be described by a Boltzmann-like equation
\be
S(V)= \lambda \log\Sigma_{\rm jammed}(V) = \lambda\log\int d\kappa\ \delta\big(V-{\cal{W}}(\kappa)\big)\Theta(\kappa) ~
\ee
where $\Theta(\kappa)$ serves to constrain the summation to reversible jammed states and $\lambda$ plays the r\^ole of Boltzmann's constant.
We may also define the analogue of temperature, compactivity
\be
X_V^{-1} = \frac{\partial S}{\partial V} ~
\ee
The partition function is then given by
\be
Z=\sum_{n}e^{-\frac{{\cal{W}}^n}{\lambda X}}\Theta_{n} ~
\ee
where the total energy is replaced by ${\cal{W}}$ and the temperature by the compactivity.

In the above setting, we may consider phase space foam as a similarly jammed system.
The packing volume corresponds to the quantum phase space volume.
The irreversible evolution corresponds to the pre-Big Bang epoch in which the Universe naturally assembles itself into a jammed state, the Big Bang singularity.
The cosmological expansion would then correspond to a reversible expansion away from the jammed state with the above defined entropy, which can indeed be shown to be an increasing function on time: $\frac{\partial S}{\partial t} \geq 0$.

The model we have presented is a generalized quantum dissipative system, {\em i.e.}, one with frictional forces at work.
Because the fluctuations of linear quantum mechanics and its associated equilibrium statistical mechanics are incapable of driving a system such as our Universe to a hot yet low entropy state and of generating a cosmological arrow of time~\cite{penrose2}, a nonlinear, non-equilibrium, strong fluctuation driven quantum theory such as the one presented here becomes necessary.
Time irreversibility is of course a hallmark of non-equilibrium systems;
this cosmological model, which notably comes with its own initial (boundary) condition, naturally produces both an arrow and an origin of time.
Moreover, in this approach the relationship of canonical quantum theory and equilibrium statistical mechanics is extended to an analogy between generalized quantum theory and non-equilibrium statistical mechanics, both of which are very much still under construction.
We will elaborate further on this interconnection~next.

\subsection{Quantum Gravity and Non-Equilibrium Physics}

We wish to discuss a more general connection between quantum gravity, the concept of holography, and some fundamental results in non-equilibrium statistical physics alluded to in the preceding sections.
This relation between holography and non-equilibrium statistical physics can prove fruitful in finding new results on both sides of the map, naturally extending the usual Wilsonian approach to quantum field theory and equilibrium statistical physics.

Let us first summarize some basic literature on non-equilibrium statistical physics.
(See~\cite{david} and references therein.)
Consider a nonlinear Hamiltonian system with slight dissipation.
The relevant nonlinearities will generate positive Lyapunov exponents in dynamics which ultimately lead to chaos;
negative Lyapunov exponents are irrelevant on long time scales.
The chaotic dynamics manifests itself in the emergence of the attractor.
There exists natural measures on this attractor, for example the Sinai--Ruelle--Bowen (SRB) measure~\cite{david}.

The {\it dissipative} dynamics is described as follows:
\begin{equation}
{\dd p_a \over \dd t} = - \kappa \frac{\partial F}{\partial p_a} ~, \qquad {\dd q_a \over \dd t} = p_a ~
\end{equation}
The dissipative term, encoded by the dissipative functional $F$, for example, goes as
\be
F \sim \sum_{ij} a_{ij} p^i p^j ~
\ee
This is generated by looking at an infinite system.
An infinite reservoir acts as a thermostat.
Integrating out the reservoir degrees of freedom produces a finite non-Hamiltonian dissipative system.
Because the dynamics is non-Hamiltonian, the volume of phase space is not preserved!
The entropy production is, at the end, crucially related to the construction of the phase space volume.

Note that it is important that there is chaos (hyperbolic dynamics) in the effective dynamical equations:
\be
\frac{\dd x}{\dd t} = g(x) ~
\ee
where the function $x(t) = f^t x(0) \sim e^{\ell t}$, with $\ell > 0$.
The positive Lyapunov exponent leads to chaotic time evolution and the stretching of some directions of a unit phase space volume.

Finally, there are measures that are invariant under time evolution.
The SRB measure $\rho$ may be singular. (It is defined on the attractor, which is usually a fractal space.) However, the time evolution averages are indeed averages over this measure, so that
\be
\lim_{T \to \infty} \frac{1}{T +t} \int_{-T}^{t} \dd\tau\ {\cal O}(f^{T +\tau} x) = \int \rho (\dd y) {\cal O}(y) ~
\label{hop}
\ee
This is the crucial relation which has a holographic interpretation. Notice that in (\ref{hop}) the dynamics of a $d+1$ dimensional system is related to an ensemble description on a $d$ dimensional system, with a specific measure. In this case the extra dimension $t$ is indeed a time parameter, but it could be some other evolution parameter such as the radial slicing of AdS space.

Another crucial aspect of this theory is that entropy production is equated with volume contraction. (Note that dissipation is crucial for the appearance of the attractor and the new measures.) The introduction of the new measures is tantamount to the breaking of the time reversal symmetry, which is ultimately the consequence of dissipation.
It is natural to ask whether this entropy production be related to the holographic entropy perhaps universally required in our formulation of quantum gravity.

The central equation for the entropy $S$ in terms of the invariant measure $\rho$ is~\cite{david}
\be
S(\rho) = \int \rho(\dd x) ( - \nabla_x \cdot g)
\ee
where the divergence is computed with respect to the phase space volume element.
For the SRB measure one can show that the $S(\rho) \ge 0$~\cite{david}.
In the light of our work it is tempting to interpret this entropy production precisely as gravitational entropy and the emergent arrow of time!
Given the naturally dynamically generated measure on the attractor, in the near-to-equilibrium limit of the response functions for this general dynamics one gets an averaged, coarse grained description as required by the ergodic theorem {\em i.e.}, an averaged ensemble description, with respect to the above measure.

We wish to point out that the non-perturbative formulation of quantum gravity reviewed in this paper is very much of the nature discussed in the context of the non-equilibrium dissipative dynamics.
Thus we would not be surprised if the well known large-scale formulation of holography, as represented by the AdS/CFT duality would be of a special type of our more general proposal.
In particular, in the case of AdS spaces, the attractor could be represented by the asymptotics of the AdS space, and the SRB distribution. This can, in special cases, be of the Gibbsian nature and may correspond to the generating functional of the dual CFT \cite{footnote3}.
It would be interesting to pursue this intuition in greater detail in the future~\cite{jarz, jarz2, jarz3}.

\section{Conclusions}

It is a widely held expectation that the theory of quantum gravity will offer a revolutionary vista on gravity and spacetime from the Planck to the Hubble scales.
Various research avenues on such a topic have appealed to a mix of rigorous mathematics, bold conjectures, foundational, and technical issues.
The foregoing presentation embodies such a blending.
Foremost, such a theory of quantum gravity must offer a solution to the problem of time.

In this paper, we have presented a model of time and of its arrow in a specific novel theory of quantum gravity.
In particular, we did so in the context of a geometric extension of standard quantum theory.
The latter, in turn, is found to be linked to a generalized non-equilibrium statistical and thermodynamic framework.
Such an extended quantum theory arises from the notion of a local, dynamical, and statistical time.
This generalizes and relaxes the global Newtonian time of quantum theory and leads to an equivalence principle, a general covariance principle, not one on spacetime, but on a dynamical space of quantum states.
The result is a harmonious nonlinear synthesis of Schr\"odinger's wave equation and the Einstein--Yang--Mills generally covariant dynamics on an infinite dimensional (almost) complex space of states.
What obtains is an emergent picture of spacetime and quantum mechanics.
How the latter structures are realized explicitly has yet to be worked out.
What we have is an informational geometric picture of quantum spacetime in which time is intrinsic and primary while space is a secondary and derivative concept.
Specifically in a scheme where dynamics is primary, time is grounded in the very nature of quantum gravitational probabilities and quantum gravitational observables within a new framework for physics.
In this context the classical Einstein paradigm about the dynamical spacetime structure has been extended to the dynamical framework for fundamental physics itself.
Moreover, the irreversibility of measurements and its consequent unresolved measurement problem in quantum theory is tied to the time arrow. That our gravity model has an arrow of time as well as specific stochastic nonlinear dynamics~\cite{weinberg} implies that it may well hold a key to the measurement problem in quantum~theory.

Clearly, as is the case with the conceptual and mathematical complexity of non-equilibrium physics, much remains to be explored in what has been laid out above.
We hope to report on our progress in future communications.

\section*{Acknowledgments}

We gratefully acknowledge stimulating discussions with Lay Nam Chang, Zach Lewis, Michel Pleimling, and Tatsu Takeuchi.
DM is supported in part by the U.S. Department of Energy under contract DE-FG05-92ER40677, task A.
The work of VJ is based upon research supported by the South African Research Chairs Initiative of the Department of Science and Technology and National Research Foundation.

\bibliographystyle{mdpi}
\makeatletter
\renewcommand\@biblabel[1]{#1. }
\makeatother

\begin{thebibliography}{100}
\bibitem{footnote1}
This extension can be also justified from a general approach to quantum correlations in the spirit of the celebrated Bell's inequality \cite{bell}.
\bibitem{footnote2}
We have presented another use of the physics of jamming in the context of the vacuum energy problem \cite{laynam, laynam2}.
\bibitem{footnote3}
Such gauge theory and string theory inspired steady state distributions have been recently proposed in the context of two-dimensional and three-dimensional turbulence~\cite{annie}.
\end{thebibliography}

\end{document}